\documentclass{pazh2col-utf}
\usepackage{graphicx}
\usepackage{amsmath}	
\usepackage{amssymb}	
\usepackage{longtable}  
\usepackage{ulem}       
\usepackage[dvipsnames]{xcolor}      
\usepackage{ragged2e}   
\usepackage{balance}    

\newcommand{\Msun}{$M_\odot$}
\newcommand{\Mmax}{M_{\text{\textsc{tov}}}}
\newcommand{\Rmax}{R_{\text{\textsc{tov}}}}
\newcommand{\csmax}{c_{\text{s\textsc{tov}}}}
\newcommand{\rhomax}{\rho_{\text{\textsc{tov}}}}
\newcommand{\Pmax}{P_{\text{\textsc{tov}}}}

\begin{document}

\authors[Ofengeim et al.]{
  \nextauth[ddofengeim@gmail.com]{D. D. Ofengeim}{1},
  \nextauth{P. S. Shternin}{2},
  \nextauth{T. Piran}{1}
}

\titles[Maximally Massive Neutron Stars]{Maximal Mass Neutron Star as a Key to Superdense Matter Physics}

\affiliations{
\nextaffil{The Herbew University of Jerusalem, Jerusalem, Israel}
\nextaffil{Ioffe Institute, Saint Petersburg, Russia}
}

\wideabstract{%
We propose a universal approximation of the equation of state of superdense matter in neutron star (NS) interiors. It contains only two parameters, the pressure and the density at the center of the maximally massive neutron star. We demonstrate the validity of this approximation for a wide range of different types of equations of state, including both baryonic and hybrid models. Combined with recently discovered correlations of internal (density, pressure, and speed of sound at the center) and external (mass, radius) properties of a maximally massive neutron star, this approximation turns out to be an effective tool for determining the equation of state of superdense matter using astrophysical observations.

\doi{10.31857/S... }

}


\section{Introduction}
\label{sec:intro}

Determining the equation of state of superdense matter is one of the central problems of neutron star astrophysics (Haensel et al. 2007). A natural approach to this problem is to infer observational constraints on the NS physical characteristics like mass $M$, radius $R$, moment of inertia $I$, etc., and then compare these with the predictions of NS structure theory (Lattimer, 2021). On this path, one usually takes a certain microphysical model, which allows one to calculate the equation of state, and by comparing the results of modeling the NS structure with observations, one then constrains the parameters of the original model.

However, the properties of matter at densities significantly larger than the saturated nuclear matter density ($\rho_0=2.8\times 10^{14}$~g/cm$^{3}$) are practically impossible to study in terrestrial laboratories. When constructing models of such matter, one has to rely on extrapolations of various methods and approaches that have proven themselves under less extreme conditions. These extrapolations are based on a variety of ideas about the microphysics of superdense nuclear matter. As a result, the ``astrophysical market'' offers a huge number of different equations of state, which at first glance are completely different from each other (see section 2 below).

It is, therefore, interesting to identify universal relations between NS parameters that weakly depend on specific microphysical models (for example, Lattimer and Prakash 2001; Bejger and Haensel 2002; Yagi and Yunes 2013a,b; Jiang and Yagi 2020; Ofengeim 2020; Cai et al., 2023). Such relations are intended to describe in a unified way the combinations of NS characteristics encountered within the framework of various approaches to modeling superdense matter. They are convenient to use for interpreting observations  as  restrictions on possible combinations of stellar parameters become largely model-independent. This work is devoted to the development of such an approach.

Many observational properties of NSs ($M$, $R$, etc.) are determined by the solution of the Tolman-Oppenheimer-Volkoff equations (Tolman, 1939; Oppenheimer and Volkoff, 1939). To close this system of equations, it is  necessary to specify the relation between pressure and density, $P=P(\rho)$. This relation, the equation of state (Haensel et al. 2007), is in the one-to-one correspondence to the $M-R$ curve of NSs (Lindblom 1992). That is if we manage to accurately determine the $M-R$ curve, the problem of finding the equation of state $P(\rho)$ will also be solved. In reality, we can only measure masses and radii of individual NSs, and only with a finite accuracy, which significantly complicates the situation.

Whatever the equation of state is, the $M-R$ curve always have a global maximum, $\Mmax$, along the mass axis (e.g. Shapiro, Teukolsky, 1985). However, the properties of the maximally massive neutron star (MMNS) are different for each equation of state model. If the latter predicts $\Mmax$ less than the mass of any of the observed NS masses, then the model is incorrect. Therefore, any observation of a sufficiently massive NS imposes significant constraints on the equation of state.

The value of $\Mmax$ specifies the natural scale of NS masses for a given equation of state. Similar characteristics are the radius of the MMNS, $\Rmax$, the density at its center, $\rhomax$, and the corresponding pressure, $\Pmax$. Note that $\rhomax$ and $\Pmax$ (for the true equation of state realized in nature) are the maximum possible density and pressure of matter in a stationary object in the contemporary Universe.  In what follows, we will also need the speed of sound at the center of the MMNS, $\csmax$.

Ofengeim (2020) showed that among the parameters characterizing the MMNS ($\Mmax$, $\Rmax$, $\rhomax$, $\Pmax$, $\csmax$), only two are independent. This was confirmed on a set of 50 equations of state for nucleonic and hyperonic compositions. Recently, a possible explanation for these findings was given in Cai et al. (2023) based on a perturbative analysis of the dimensionless Tolman-Oppenheimer-Volkoff equations.
In the present work, the existence of correlations between these parameters is confirmed on an extended sample of 162 equations of state, including nucleon, hyperon, and hybrid models (i.e., with a quark inner core), and new compact approximation formulas are proposed to describe these correlations.

In addition, we constructed a universal approximation of the $P(\rho)$ relations for $\rho\gtrsim 3\rho_0$ which is based on only two parameters, namely $\rhomax$ and $\Pmax$. Taking into account the one-to-one correspondence between the pair $\rhomax,\Pmax$   and the pair $\Mmax,\Rmax$, this provides a model-independent tool for a direct transformation of observational constrains on the properties of MMNS into constraints on the equation of state of superdense matter at densities that are most difficult to reach in laboratory research.

\section{The Zoo of equations of state}
\label{sec:zoo}
\begin{figure*}
    \includegraphics[width=0.325\textwidth]{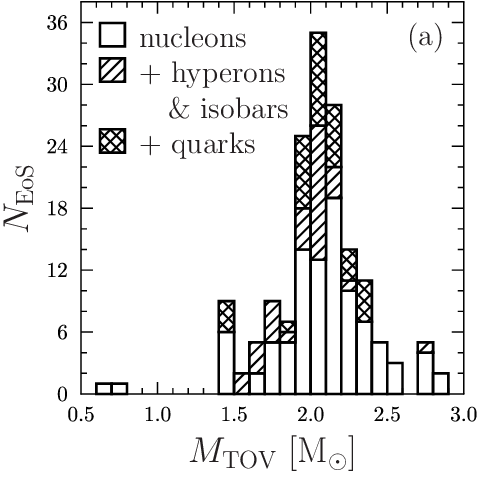}
    \includegraphics[width=0.325\textwidth]{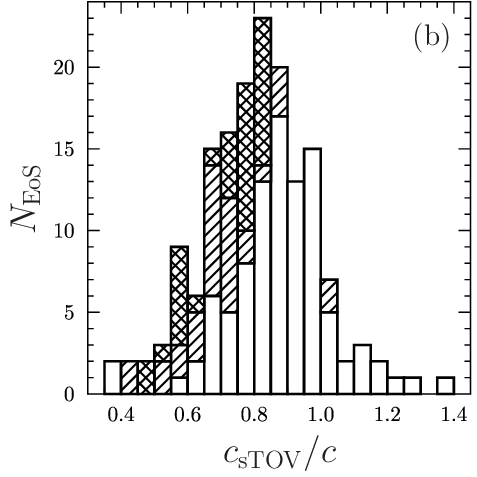}
    \hspace{0.003\textwidth}
    \includegraphics[width=0.325\textwidth]{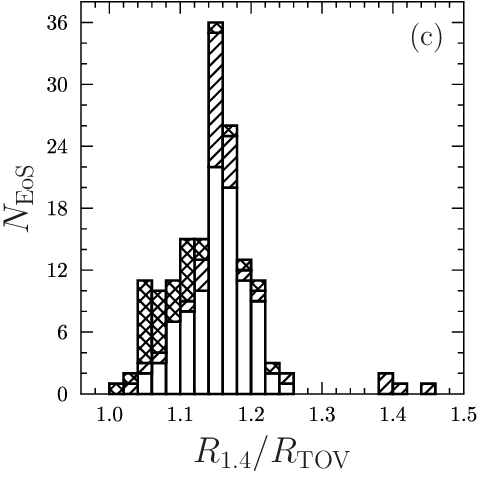}
    \caption{\justifying 
    Distributions  of (a) NS maximal mass, (b) speed of sound in the center of the star, (c) $R_{1.4}/\Rmax$ ratio for the  sample of equations of state used. Different hatching patterns corresponds to different model types (nucleonic, hyperonic, hybrid).
    }
    \label{fig:eosStat}
\end{figure*}
We consider 162 NS equations of state. Among them, 97 have nucleonic composition, 32 allow for the appearance of hyperons and $\Delta$-isobars (but do not allow for free quarks) in NS interiors, and 33 equations of state predict the presence of internal quark core of the star. Main sources for equations of state are the CompOSE database\footnote{
\texttt{https://compose.obspm.fr.}
} (Typel et al., 2015), the collection of models from Read et al. (2009) and models used earlier by Ofengeim (2020). A complete list of the equations of state used containing heir main characteristics and references is given in Appendix.

The considered models are based on a variety of approaches to modeling nuclear interactions and the microphysics of superdense matter. These include obsolete models of free degenerate neutron gas and free $npe$ gas; models based on effective energy density functionals, including purely phenomenological (PAL, PAPAL, BGN families), non-relativistic Skyrme-type (SLy, BSk, SkI, etc.) or Gogni (D1M*) and others, numerous relativistic mean field functionals; models derived from microscopic baryon interaction potentials using many-body methods (APR, WFF, BBB); and also some other models. Among the hybrid equations of state, both those in which a first-order phase transition occurs between hadronic and quark matter (for example, CMF, VQCD) and those in which there is a quark-hadron crossover (QHC) between these phases are present.

Figure~\ref{fig:eosStat} shows how this sample of models is distributed over $\Mmax$ (panel a), over $\csmax$ (panel b), and over the ratio of the radius $R_{1.4}$ of a ``canonical'' NS with a mass of $1.4$\Msun\ to the radius of the most massive star, $\Rmax$ (panel c). Notice that all 162 equations of state satisfy the condition $\Rmax<R_{1.4}$.

Realistic equations of state, apparently, should satisfy the causality condition $\csmax < c$, where $c$ is the speed of light in vacuum (Haensel et al. 2007), and explain present observations of massive radio pulsars (Demorest et al., 2010; Antoniadis et al., 2013; Fonseca et al., 2021). About one third of the equations of state of our sample violate the second condition, and about 10\% violate the first one. We do believe, however, that it is important to include both realistic and non-realistic models in order to comprehensively explore the generality of the detected correlations and the constructed approximations.

We also note, that the speed of sound is not always a monotonic function of the density (especially for non-nucleonic models), and $\csmax$ is not necessarily the highest possible speed of sound within a star. However, it turns out that the more stringent condition, $c_\text{s} < c$ in the entire volume of the star, is violated by the same number of equations of state as the condition $\csmax<c$.

\begin{figure*}
\centering
    \includegraphics[width=\textwidth]{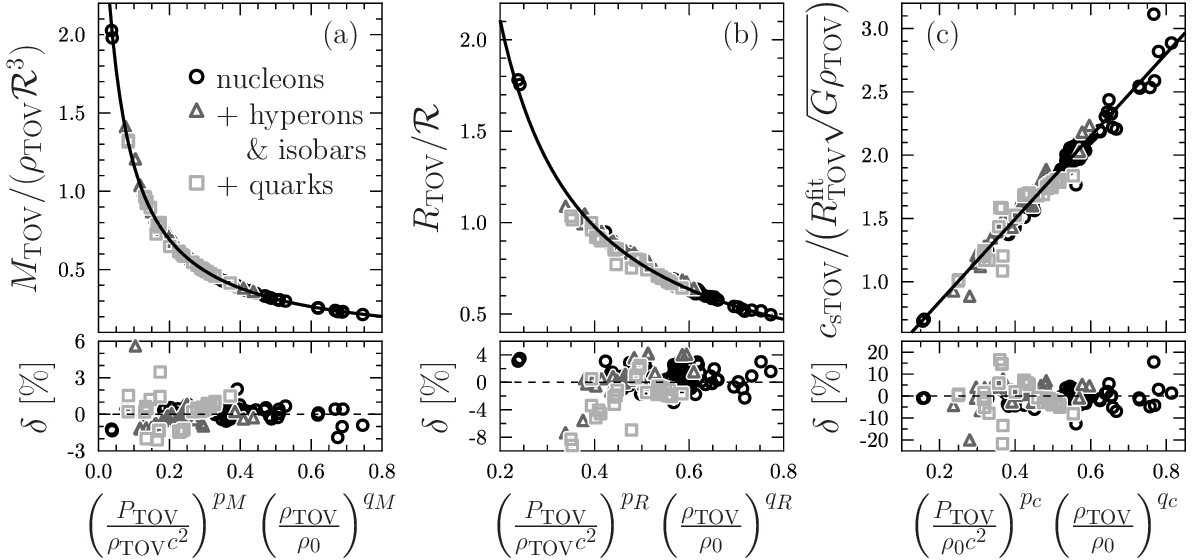}
\caption{\label{fig:corrs}\justifying Correlations of MMNS characteristics for different equations of state. Different symbols correspond to different models ($\pmb{\bigcirc}$ to nucleonic models,  $\pmb{\triangle}$ to hyperonic and isobar models,  $\pmb{\Box}$ to hybrid models, respectively). The solid lines show the approximations given in equations~(\ref{eq:maxCorrs}). The quantity $\mathcal{R}$ is defined in equation~(\ref{eq:maxCorrs-calR}), $R_{\rm TOV}^\text{(fit)}$ is given by equation~(\ref{eq:maxCorrs-R}). The exponents $p_i$ and $q_i$, with $i = M,R,c$, are given in table~\ref{tab:maxCorrCoeffs}. Relative errors of the fits are shown in lower panels.}
\end{figure*}

\section{Correlations of MMNS characteristics}
\label{sec:maxNS}
There are strong correlations between the values of $\Mmax$, $\Rmax$, $\rhomax$, $\Pmax$, and $\csmax$ calculated within different models (Ofengeim, 2020; Cai et al., 2023), for which we propose here new fitting formulae:
\begin{subequations}
\label{eq:maxCorrs}
\begin{equation}
    \label{eq:maxCorrs-M}
    \Mmax = \frac{\rhomax \mathcal{R}^3}{f_M(\Pmax,\rhomax)},
\end{equation}
\begin{equation}
    \label{eq:maxCorrs-R}
    \Rmax = \frac{\mathcal{R}}{f_R(\Pmax,\rhomax)}, 
\end{equation}
\begin{equation}
     \label{eq:maxCorrs-c}
     \csmax =\sqrt{G\rhomax}\ \mathcal{R}\ \frac{f_c(\Pmax,\rhomax)}{f_R(\Pmax,\rhomax)}. 
\end{equation}
These formulas differ from those proposed earlier by Ofengeim (2020) and Cai et al. (2023) and are based on dimensional analysis. The key to their construction is the introduction of the characteristic ``Jeans'' scale for the radius
\begin{equation}
    \label{eq:maxCorrs-calR}
    \mathcal{R}=\sqrt{\frac{\Pmax}{G\rhomax^2}}
\end{equation}
and the corresponding scale for the mass  $\rhomax \mathcal{R}^3$. The dimensionless functions $f_M$, $f_R$ and $f_c$ have the same structure
\begin{equation}
    \label{eq:maxCorrs-f}
    f_i = c_i \left(\frac{\Pmax}{\rhomax c^2}\right)^{p_i} \left(\frac{\rhomax}{\rho_0}\right)^{q_i} + d_i,
\end{equation}
\end{subequations}
where the optimal values of the fitting parameters $c_i$, $d_i$, $p_i$ and $q_i$ are given in table~\ref{tab:maxCorrCoeffs}. Figure~\ref{fig:corrs} clearly demonstrates the accuracy of the obtained approximations. It can be seen that the correlations of $\Mmax$ and $\Rmax$ with $\Pmax$, $\rhomax$ are described with good accuracy, while the approximation formula for $\csmax$ shows a larger scatter.

\begin{table}
 \centering\small
	\caption{\label{tab:maxCorrCoeffs}\justifying Parameters of the approximations given in equation~(\ref{eq:maxCorrs}). The last two columns show the relative root mean squared fit error  (rrms) and the maximal relative fit error  (max). Models-outliers JJ(VQCD)soft and JJ(VQCD)intermediate are not included in the calculation of these errors.}
   \renewcommand{\arraystretch}{1.4}
   \setlength{\tabcolsep}{0.015\columnwidth}
	\begin{tabular}{lcccccc}
		\hline\hline
		$i$  & $p_i$     & $q_i$     & $c_i$    & $d_i$      & rrms    & max\\
		\hline
		$M$  & $1.41$  & $0.0177$ & $5.86$ & $0.273$  & 0.86\%  & 5.4\%    \\
        $R$  & $0.518$  & $-0.0755$ & $2.74$  & $-0.0741$ & 2.0\%   & 9.3\%    \\
        $c$  & $0.76$   & $-0.016$  & $3.27$   & $0.19$    & 4.9\%   & 22\%     \\
		\hline\hline
	\end{tabular}
\end{table}

Among the 162 models considered, there are two that stand out significantly from these correlations. These are the JJ(VQCD)soft and JJ(VQCD)intermediate models, for which the center of the most massive NS is at the verge of the region of the first-order phase transition. In this case, the maximum of the $M-R$ curve turns out to be non-smooth, which affects the correlation $\Mmax(\rhomax,\Pmax)$. These models are not shown in the bottom panels of Figure~\ref{fig:corrs}. In all other cases, when the maximum of the $M-R$ curve is smooth, no significant deviations from correlations~(\ref{eq:maxCorrs}) are observed.

\section{The universal approximation  $P(\rho)$}
\label{sec:P-rho}

If we express the dependencies $P(\rho)$ in dimensionless variables $P/\Pmax$, $\rho/\rhomax$, the behavior of all 162 models will be very similar, especially when approaching the center of the MMNS.  A good universal approximation  at $\rho>3\rho_0$ is: 
\begin{subequations}\label{eq:Prho}
\begin{equation}\label{eq:Prho-P}
    P = \Pmax\ g_P\left( \rho/\rhomax, \csmax^\text{(fit)}/c, \gamma_\text{max}^\text{(fit)} \right) \ , 
\end{equation}
where $\csmax^\text{(fit)}$ is given by  equation~(\ref{eq:maxCorrs-c}), the value $\gamma_\text{max}^\text{(fit)} = \left(\csmax^\text{(fit)}\right)^2 \rhomax/\Pmax$ plays the role of the adiabatic index\footnote{
$\gamma_\text{max}^\text{(fit)}$ may differ significantly from the actual value of the adiabatic index at the point $\rho=\rhomax$. Although the fitting error~(\ref{eq:maxCorrs-c}) for $\csmax$ is generally not too large, but the error of $\csmax^2$ in a few worst cases exceeds $50\%$.
} in the MMNS center, and the function $g_P$ is
\begin{multline}
    \label{eq:Prho-g}
    g_P(x,\zeta,\gamma) = x^{\gamma-a_0-a_1\zeta}\Bigl[ 1\: +\\
    +(a_0+a_1\zeta)(1-x)+(b_0+b_1\zeta)(1-x)^p \Bigr]^{-1}
\end{multline}
with 
$a_0 = -0.6268$, 
$a_1 = -0.1294$,
$b_0 = -0.5588$,
$b_1 = 1.023$ 
and
$p = 2.494$. 
\end{subequations}

\begin{figure}[!t]
    \includegraphics[width=\columnwidth]{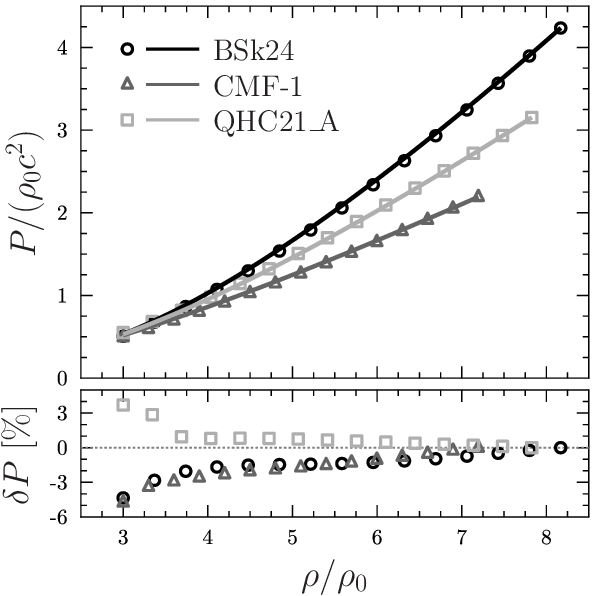}
    \caption{\label{fig:ExamplesPrho}\justifying Typical examples of the application of the approximation ~(\ref{eq:Prho}) (lines) to nucleonic (BSk24, circles), hyperonic (CMF-1,traingles) and hybrid (QHC21\_A, squares) equations of state. Each curve is plotted in the region $3\rho_0\leqslant \rho \leqslant \rhomax$. The lower panel shows the relative approximation errors.}
\end{figure}
On average, the approximation error~(\ref{eq:Prho}) does not exceed a few percent. Figure~\ref{fig:ExamplesPrho} shows three typical examples for the nucleon, hyperon, and hybrid equations of state. It can be seen that the fit works best at high densities, and is worst at lower ones.
\begin{figure}[!t]
    \includegraphics[width=\columnwidth]{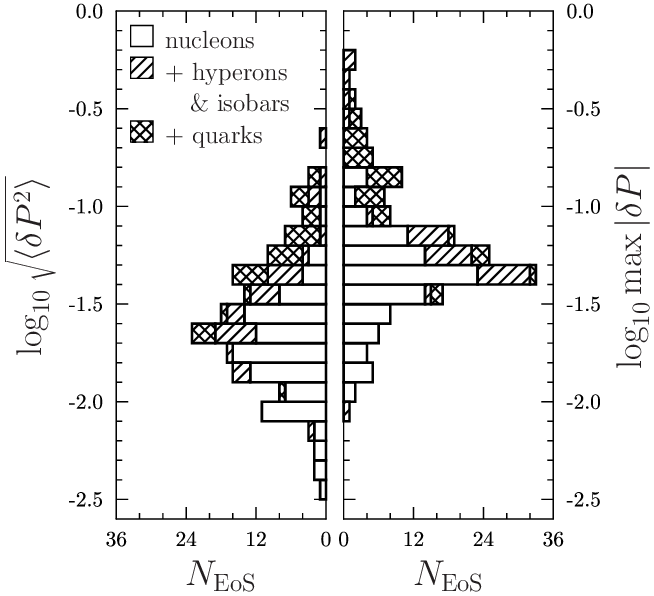}
    \caption{\label{fig:ErrHistPrho}\justifying The distribution of the errors of the fit~(\ref{eq:Prho}) over the  whole sample of equations of state.}
\end{figure}

Figure~\ref{fig:ErrHistPrho} shows the distributions of the root-mean-square and maximal deviations from the approximation~(\ref{eq:Prho}) at $\rho>3\rho_0$ for the  sample of equations of state (see also the table~\ref{tab:zooNuc} in Appendix). For nucleonic models, even the largest error does not exceed 10\%. The error can be significantly higher  for models with phase transitions (either to hyperonic or quark matter). About 10 equations of state give a maximal error of more than $30-60\%$, but for most of them the standard deviation is still $\lesssim 10\%$. This is a manifestation of the heteroscedasticity of the fitting error, already demonstrated in Figure~\ref{fig:ExamplesPrho}. The standard deviation is dominated by the region of large densities, where the approximation works well, while the largest error, on the contrary, characterizes the fit quality at low densities. 

Only one model, RSGMT(QMC700), gives a large, $\sim 20\%$, root mean square error, i.e. it is unsatisfactorily described by proposed approximations at any density. However, it has extreme values even for those properties that are subject to laboratory testing, and therefore seems unrealistic.

\begin{figure*}[!t]
    \centering
    \includegraphics[width=\textwidth]{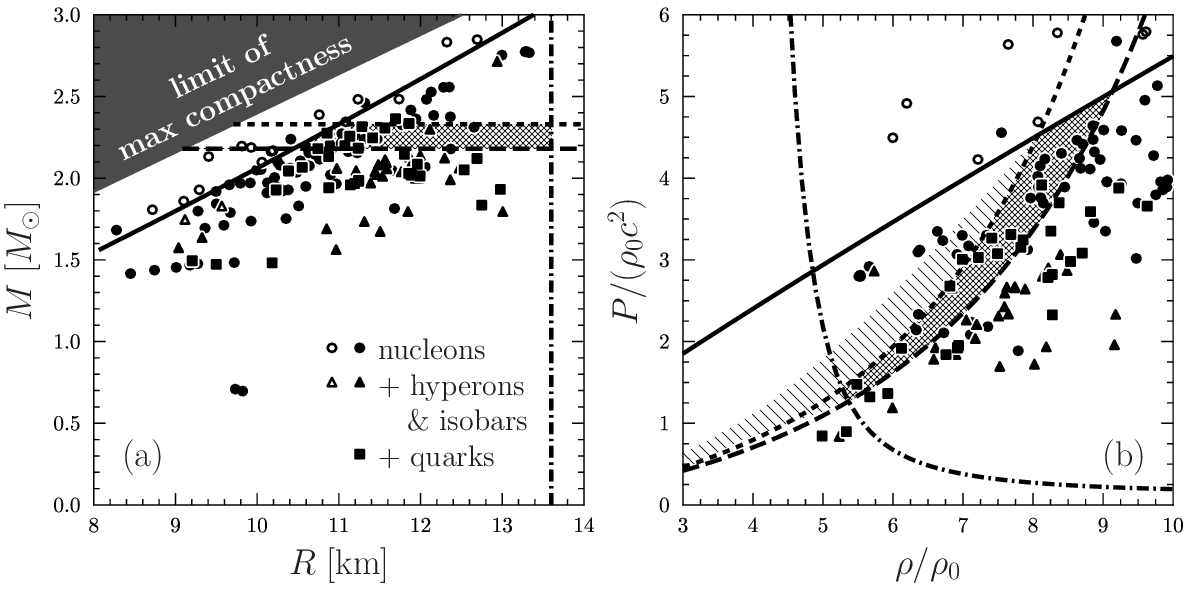}
    \caption{\label{fig:double}\justifying  MMNSs positions on  $M,R$ (a) and $P,\rho$ (b) planes. Filled symbols show subluminal equations of state ($\csmax<c$), open symbols correspond to superluminal ones ($\csmax\geqslant c$). Solid lines show the boundaries $\csmax=c$, obtained using approximations~(\ref{eq:maxCorrs}). Long-dashed lines show the constraint from Kandel and Romani (2023),  $\Mmax>2.19$\Msun, short-dashed lines show the constraint $\Mmax < 2.33$\Msun\ from Rezzolla et al. (2018). Dash-dotted line shows the constraint $\Rmax < R_{1.4} < 13.6\,$km (Annala et al., 2018). Regions which fulfill all these constraints are doubly hatched. The single-hatched region in panel (b) shows the region occupied by all $P(\rho)$ curves, whose MMNS points $(\rhomax,\Pmax)$  are anywhere  in the allowed (double-hatched) region. This region describes the allowed equations of state under the constraints imposed here.
    Note that for readability the scale of (b) was chosen such that  some of the models are out of the boundaries of the figure.}
\end{figure*}

\section{Applications to observational constraints on MMNSs}
\label{sec:obs}

We turn now to demonstrate how the proposed universal correlations enable us to impose model-independent constraints on the equation of state of superdense matter based on observational data.

Figure~\ref{fig:double}(a) show MMNS positions on the $M,R$ plane for all considered equations of state (i.e., each symbol is a point $(\Rmax,\,\Mmax)$) corresponding to a specific equation of state. Different types of symbols correspond to different types of equations of state (nucleonic, hyperonic, hybrid). In addition, open symbols indicate ``superluminal'' models for which $\csmax\geqslant c$. The shaded area in the upper left corner shows the so-called maximum compactness limit for possible values of NS masses and radii (for example, Lattimer and Prakash, 2016). Figure~\ref{fig:double}(b) shows the same data, but on the $P,\rho$ plane (each symbol is a point $(\rhomax,\,\Pmax)$), however for the sake of readability, a scale was chosen so that some of the models fell out of the boundaries of the figure.

There are various constraints on $\Mmax$ in the literature with varying degrees of certainty. The maximal mass is limited from below by observations of the most massive NSs. As an example of such a restriction, we chose the condition from Kandel and Romani (2023), namely $\Mmax > 2.19 M_\odot$ at the 2$\sigma$ significance level. It is based on measurements of the masses of pulsars in binary systems. This limit is shown in Figure~\ref{fig:double}(a) with the long-dashed line. In addition, Rezzolla et al. (2018), based on an analysis of the neutron star merger event GW170817 (Abbott et al., 2017), proposed an upper limit on the mass of the MMNS, $\Mmax<2.33 M_\odot$ at the 2$\sigma$ significance level. This limit is shown in Figure~\ref{fig:double}(a) with the short-dashed line. In the following we use these constraints for illustrative purposes and demonstrate how observational constraints can be used to put limits on the equation of state.

Direct inference of the MMNS radius from observations is difficult. We will resort to the following considerations. As shown in Figure~\ref{fig:eosStat}(c), for all 162 equations of state $\Rmax$ turns out to be less than the radius $R_{1.4}$ of the ``canonical'' NS  with $M=1.4$\Msun. Inferences of the constraints on the radii of medium-mass NSs are simpler due to their greater accessibility for various types of observations (for example, Degenaar and Suleimanov, 2018). Conservatively, we employ the results of the analysis of the gravitational wave event GW170817 from the work of Annala et al. (2018), in which $R_{1.4}<13.6\,$km was obtained at the $90\%$ significance level\footnote{Stronger limits can be obtained from other observations}. 
We set the same condition on $\Rmax$, which is shown in Figure~\ref{fig:double}(a) with the dash-dotted line.

These three constraints can be mapped onto the $P,\rho$ plane by considering the fits (\ref{eq:maxCorrs-M}) and (\ref{eq:maxCorrs-R}) as equations for determining $\rhomax$ and $\Pmax $. Their solutions are depicted in Figure~\ref{fig:double}(b) with lines of the same styles as the corresponding constraints in Figure~\ref{fig:double}(a).

Finally, realistic equations of state must satisfy the theoretical restriction $\csmax<c$. Using formulas~(\ref{eq:maxCorrs}), this boundary is shown on both planes of figure~\ref{fig:double} with the solid lines. The true MMNS should be located below these lines. Notice that due to the approximate nature of the fitting formulas, a small number of superluminal models (empty symbols) fall below, and a small number of subluminal (filled symbols) fall above the solid lines. This uncertainty, as well as the uncertainties in transferring observational constraints from $\Mmax$ and $\Rmax$ to $\Pmax$ and $\rhomax$, must be taken into account in a thorough analysis, but here we neglect them in our illustrative analysis.

Combining these constraints, we obtain the allowed regions of $(\Mmax,\,\Rmax)$ and $(\Pmax,\,\rhomax)$ which are shown in figure~\ref{fig:double} as a double hatch region. Using approximation~(\ref{eq:Prho}) from each point of this region one can draw a curve towards lower densities. The set of these curves will then occupy the area shown in Figure~\ref{fig:double}(b) as a single hatch region. This area represents the final constraint on the equation of state that is constructed using this method. It appears to be at least as stringent as  limits proposed in other recent work (Raaijmakers et al., 2021; Jiang et al., 2023). But note that this result is purely illustrative. A more thorough analysis requires taking into account uncertainties in the approximations (\ref{eq:maxCorrs}) and (\ref{eq:Prho}) and analyzing the systematic errors of the employed observations, as well as accounting for other observations that we haven't used here.

\section{Conclusions}
\label{sec:concl}

We elaborated and improved the method for determining the equation of state of superdense matter from the properties of MMNS, previously proposed by Ofengeim (2020). The correlations between the quantities $\Mmax$, $\Rmax$, $\Pmax$, $\rhomax$ and $\csmax$ discovered in that work are confirmed on a much wider set of equations of state, which now includes both baryonic and hybrid models. Using dimensional analysis we present new approximate formulae~(\ref{eq:maxCorrs}) that describe theses correlations.

In addition, it is shown that the $P(\rho)$ relations for a variety of the equation of state models can be described by a single approximation~(\ref{eq:Prho}), which has only two parameters, namely pressure $\Pmax$ and density $\rhomax$ at the center of MMNS. This fit has a low accuracy at small densities $\rho\lesssim 3\rho_0$, but it works well in the more interesting denser region.

By applying formulas~(\ref{eq:maxCorrs}) to a number of theoretical and observational constraints on the characteristics of MMNS, we limit significantly the possible values of $\Mmax$, $\Rmax$, $\Pmax$ and $\rhomax$. Using the formula~(\ref{eq:Prho}), the constraints on the last pair  are converted into constraints on the entire equation of state in the region $\rho\gtrsim 3\rho_0$.

The nature of the universal properties studied here remains unclear. In the work by Ofengeim~(2020), the existence of correlations between the properties of MMNS was interpreted as indirect confirmation of the approximate two-parametric nature of realistic equations of state identified by Lindblom (2010). The universality of the $P(\rho)$ curves reported in the present work, at first glance, directly extends Lindblom’s findings to almost all existing models and proposes a pair of quantities $\Pmax$ and $\rhomax$ (or, taking into account correlations~(\ref{eq:maxCorrs}), $\Mmax$ and $\Rmax$) as these two parameters. However, Lindblom's two-parametric description had an ``anchor'' point at the crust-core boundary, i.e. at a density of $\sim 0.5\rho_0$, while the expression~(\ref{eq:Prho}) works above $3\rho_0$. Therefore, the parametrization of the equations of state with two parameters used in this work is somewhat different from that discovered by Lindblom (2010). In addition, both the Lindblom parametrization and the parametrizations of the $P(\rho)$ curves proposed here are completely phenomenological and have no microscopic justification. Finally, Cai et al. (2023) recently proposed an explanation for the correlations between the characteristics of the MMNS, which does not rely at all on the properties of the equation of state. So perhaps the correlations~(\ref{eq:maxCorrs}) and the approximations~(\ref{eq:Prho}) have different physical origins.

This work can be extended in three natural directions. First, one can extrapolate the approximation of $P(\rho)$ curves to the region $\rho < 3\rho_0$. In order to do so, it probably will be necessary to increase the number of independent parameters in the fitting formula. Second, due to the one-to-one correspondence between the relations $P(\rho)$ and $M,R$ (Lindblom, 1992), the latter set of curves should also be describable by a small number of real parameters. Accordingly, one can try to propose a universal approximation for the $M,R$ curves. Such approximation will be useful in almost all studies related to observations of NS masses and radii. Finally, it is necessary to determine which of the explanations for the correlations of MMNS properties, those of Ofengeim (2020), Cai et al. (2023) or some other is correct. Regardless of these open questions, we expect that the method presented here can be used to infer new limits from observations on the equation of state of matter at very high densities.

\vspace{1ex}
The work was supported by RSF grant \# 19-12-00133 (PS) and by an Advanced ERC grant MultiJets (DO, TP).


\appendix
\onecolumn
\section*{Supplementary Materials: Equations of state zoo catalogue}

The selection of equations of state for this work was carried out as follows. From the array of CompOSE models (Typel et al., 2015) (as of November 2022), the category \texttt{Cold Neutron Star EoS} was taken. Only the tables covering both the NS crust and core were used. Models from Read et al. (2009) were added to this sample (this collection of equations of state was available for some time through the link provided in the review by Özel and Freire, 2016; an archived copy is available now), and several dozen models more which are available to the authors (see Ofengeim, 2020). Then duplicates and equations of state tables of which ended at $\rho<\rhomax$ were discarded. As a result, we selecteed a sample of 162 models, presented in the table~\ref{tab:zooNuc}.

The nomenclature used here follows the CompOSE tradition and is based on the following principle: \textit{<acronym of the original paper>(<model designation>)}. Tables for each equation of state, as well as its detailed description (nuances of choosing microphysical parameters, crustal models, etc.) can be found in the references indicated in the ``source'' column. There one can also find a decoding of the acronym of the original work where this particular model was proposed (which rarely coincides with the source). If neither the source nor the references given there describe a model for a crust, it means that the BSk24 crust was used (Pearson et al., 2018), stitched to the core by requiring the continuity of pressure at the boundary. Model designations in the table~\ref{tab:zooNuc} may undergo reasonable abbreviations in comparison with those used in the referenced sources.

Additionally, for each equation of state the maximum mass of the NS, its radius, pressure, density and speed of sound at the center of such a star, as well as the radius of the <<canonical>> NS $R_{1.4}$ are indicated in the table~\ref{tab:zooNuc}. The root mean square and maximum relative fitting errors of the $P(\rho)$ fit~(2) are also indicated for the range $\rho>3\rho_0$.

{\small
\renewcommand{\arraystretch}{1.4}
\setcounter{table}{0}
\renewcommand{\thetable}{S\arabic{table}}
\begin{longtable}{l|c|ccc|ccc|cc}
    \caption{\label{tab:zooNuc} Neutron star equations of state \vspace{-2ex}}\\
    \hline\hline
    Name & Reference & $\Mmax$ & $\Rmax$ & $R_{1.4}$ & $\rhomax$  & $\Pmax$        & $\csmax$ & \multicolumn{2}{c}{Fit $P(\rho)$ [\%]}  \\
             &          & [\Msun] & [km]    & [km]      & $[\rho_0]$ & $[\rho_0 c^2]$ & $[s]$    & $\sqrt{\langle\delta P^2\rangle}$                 & $\max |\delta P|$                                   \\
    \hline \endfirsthead 
    \multicolumn{10}{l}{\textit{Continued}}\\
    \hline
    Name & Reference & $\Mmax$ & $\Rmax$ & $R_{1.4}$ & $\rhomax$  & $\Pmax$        & $\csmax$ & \multicolumn{2}{c}{Fit $P(\rho)$ [\%]}  \\
             &          & [\Msun] & [km]    & [km]      & $[\rho_0]$ & $[\rho_0 c^2]$ & $[s]$    & $\sqrt{\langle\delta P^2\rangle}$                 & $\max |\delta P|$                                   \\
    \hline \endhead 
    \hline\multicolumn{10}{l}{\textit{Continues on the next page}}\\
    \endfoot
    \hline\hline 
    \multicolumn{10}{c}{\parbox{0.95\textwidth}{\vspace*{1ex}
References: R+09 --- Read et al. (2009), \"{O}zel and Freire (2016), available through the archive url \texttt{https://web.archive.org/web/20200830055909/http://xtreme.as.arizona.edu/neutronstars/}; HPY07 --- Haensel et al. (2007); G+05 --- Gusakov et al. (2005); K+14 --- Kaminker et al. (2014); O20 --- Ofengeim (2020); Po+13 --- Potekhin et al. (2013); Pe+18 --- Pearson et al. (2018); Y+11 --- Yakovlev et al. (2011); O+19 --- Ofengeim al. (2019); GHK14 --- Gusakov et al. (2014); CompOSE --- CompStar Online Supernova Equations of State database, \texttt{https://compose.obspm.fr} (Typel et al., 2015).
}}
    \endlastfoot
    \multicolumn{10}{c}{Nucleonic models}\\
    APR(APR1) & R+09 & 1.68 & 8.28 & 9.34 & 14.9 & 7.42 & 0.839 & 8.6 & 12 \\
APR(APR2) & R+09 & 1.81 & 8.72 & 10.16 & 13.4 & 7.63 & 1.03 & 5.4 & 9.8 \\
APR(APR3) & R+09 & 2.39 & 10.76 & 12.06 & 8.35 & 5.78 & 1.13 & 1.9 & 7.5 \\
ZBL(BPAL12) & R+09 & 1.45 & 9.01 & 10.04 & 14.2 & 4.29 & 0.704 & 2.8 & 9.3 \\
EOHJBO(ENG) & R+09 & 2.24 & 10.41 & 12.01 & 9.19 & 5.68 & 1.00 & 3.5 & 6.7 \\
FP(FPS) & R+09 & 1.80 & 9.28 & 10.84 & 12.1 & 5.40 & 0.870 & 2.7 & 5.4 \\
PAL(PAL6) & R+09 & 1.48 & 9.27 & 10.54 & 13.5 & 3.93 & 0.692 & 1.7 & 5.9 \\
WFF(WFF1) & R+09 & 2.13 & 9.41 & 10.39 & 10.8 & 8.04 & 1.18 & 6.2 & 16 \\
WFF(WFF2) & R+09 & 2.20 & 9.81 & 11.14 & 10.1 & 7.37 & 1.15 & 2.9 & 4.9 \\
WFF(WFF3) & R+09 & 1.84 & 9.50 & 10.90 & 11.3 & 4.86 & 0.836 & 2.8 & 4.6 \\
MPA(MPA1) & R+09 & 2.46 & 11.32 & 12.44 & 7.55 & 4.56 & 0.991 & 2.6 & 7.0 \\
MS(MS1) & R+09 & 2.77 & 13.33 & 14.84 & 5.53 & 2.80 & 0.892 & 0.94 & 1.4 \\
MS(MS1b) & R+09 & 2.78 & 13.29 & 14.52 & 5.51 & 2.80 & 0.894 & 0.95 & 1.4 \\
MS(MS2) & R+09 & 1.81 & 11.68 & 13.72 & 7.79 & 1.89 & 0.582 & 2.3 & 7.2 \\
ABHT(QMC-RMF1) & CompOSE & 1.95 & 10.25 & 11.85 & 9.82 & 3.96 & 0.806 & 1.5 & 4.6 \\
ABHT(QMC-RMF2) & CompOSE & 2.04 & 10.49 & 12.02 & 9.26 & 3.93 & 0.828 & 1.6 & 5.0 \\
ABHT(QMC-RMF3) & CompOSE & 2.14 & 10.71 & 12.24 & 8.81 & 4.11 & 0.862 & 2.5 & 7.2 \\
ABHT(QMC-RMF4) & CompOSE & 2.21 & 11.03 & 12.33 & 8.15 & 3.69 & 0.852 & 1.6 & 5.2 \\
APR(APR) & CompOSE & 2.19 & 9.93 & 11.33 & 9.93 & 6.87 & 1.15 & 1.8 & 6.0 \\
BG(BGN1) & HPY07 & 2.17 & 10.91 & 13.00 & 8.66 & 4.24 & 0.939 & 2.4 & 4.1 \\
BG(BGN2) & HPY07 & 2.48 & 11.74 & 13.53 & 7.21 & 4.23 & 1.04 & 3.3 & 6.4 \\
BBB(BBB1) & HPY07 & 1.79 & 9.67 & 11.03 & 11.0 & 4.05 & 0.765 & 2.9 & 6.4 \\
BBB(BHF-BBB2) & CompOSE & 1.92 & 9.50 & 11.12 & 11.4 & 5.75 & 0.915 & 2.4 & 3.6 \\
BL(chiral) & CompOSE & 2.08 & 10.28 & 12.31 & 9.78 & 5.13 & 0.985 & 1.2 & 1.7 \\
GDTB(DDH$\delta$) & CompOSE & 2.15 & 11.23 & 12.65 & 7.92 & 3.13 & 0.790 & 1.3 & 4.9 \\
GKYG(APR-I) & G+05 & 1.92 & 10.31 & 12.14 & 9.85 & 3.86 & 0.801 & 0.83 & 3.7 \\
GKYG(APR-II) & G+05 & 1.92 & 10.27 & 12.07 & 9.91 & 3.89 & 0.802 & 0.81 & 3.7 \\
GKYG(APR-III) & G+05 & 1.93 & 10.38 & 12.26 & 9.75 & 3.80 & 0.799 & 0.92 & 3.8 \\
GM(GM1) & CompOSE & 2.38 & 12.16 & 14.19 & 6.91 & 3.07 & 0.848 & 2.2 & 6.3 \\
GMSR(BSk14) & CompOSE & 1.93 & 9.29 & 11.17 & 11.9 & 7.10 & 1.13 & 4.9 & 14 \\
GMSR(H7) & CompOSE & 2.48 & 11.23 & 12.68 & 7.64 & 5.64 & 1.38 & 10 & 15 \\
GMSR(LNS5) & CompOSE & 1.97 & 9.92 & 11.48 & 10.4 & 4.84 & 0.890 & 2.2 & 5.8 \\
GMSR(SLy5) & CompOSE & 2.10 & 10.06 & 11.79 & 10.0 & 5.67 & 1.00 & 0.81 & 2.0 \\
GPPVA(DD2) & CompOSE & 2.42 & 11.88 & 13.19 & 6.99 & 3.30 & 0.866 & 1.3 & 3.9 \\
GPPVA(DDME2) & CompOSE & 2.48 & 12.07 & 13.23 & 6.71 & 3.24 & 0.873 & 1.4 & 2.9 \\
GPPVA(FSU2) & CompOSE & 2.07 & 12.06 & 13.89 & 7.11 & 2.08 & 0.632 & 1.7 & 4.3 \\
GPPVA(FSU2H) & CompOSE & 2.38 & 12.36 & 13.30 & 6.36 & 2.33 & 0.703 & 3.5 & 5.0 \\
GPPVA(FSU2R) & CompOSE & 2.05 & 11.65 & 12.95 & 7.35 & 2.18 & 0.632 & 2.5 & 3.7 \\
GPPVA(NL3$\omega\rho$L55) & CompOSE & 2.75 & 13.00 & 13.73 & 5.66 & 2.92 & 0.894 & 1.8 & 2.6 \\
GPPVA(TM1e) & CompOSE & 2.12 & 11.84 & 13.17 & 7.12 & 2.24 & 0.654 & 2.3 & 3.3 \\
GPPVA(TW) & CompOSE & 2.08 & 10.64 & 12.33 & 9.05 & 3.95 & 0.839 & 1.8 & 6.3 \\
KKPY(APR-IV) & K+14 & 2.16 & 10.86 & 12.72 & 8.68 & 4.12 & 0.904 & 1.6 & 4.8 \\
free $n$ & O20 & 0.709 & 9.73 & --- & 15.0 & 1.44 & 0.369 & 0.58 & 1.2 \\
free $npe$ & O20 & 0.697 & 9.82 & --- & 14.7 & 1.36 & 0.362 & 0.69 & 1.5 \\
PCGS(PCSB0) & CompOSE & 2.53 & 12.13 & 13.30 & 6.63 & 3.35 & 0.888 & 2.2 & 3.2 \\
PFCPG(BSk19) & Po+13 & 1.86 & 9.10 & 10.73 & 12.4 & 6.76 & 1.01 & 3.9 & 8.7 \\
PFCPG(BSk20) & Po+13 & 2.16 & 10.17 & 11.74 & 9.61 & 5.79 & 1.06 & 0.53 & 2.6 \\
PFCPG(BSk21) & Po+13 & 2.27 & 11.04 & 12.59 & 8.17 & 4.24 & 0.963 & 1.6 & 4.4 \\
PCPFDDG(BSk22) & Pe+18 & 2.26 & 11.20 & 13.04 & 8.07 & 4.02 & 0.951 & 2.6 & 4.6 \\
PCPFDDG(BSk24) & Pe+18 & 2.28 & 11.07 & 12.57 & 8.10 & 4.15 & 0.955 & 1.4 & 4.8 \\
PCPFDDG(BSk25) & Pe+18 & 2.22 & 11.04 & 12.36 & 8.11 & 3.76 & 0.896 & 0.90 & 4.1 \\
PCPFDDG(BSk26) & Pe+18 & 2.17 & 10.19 & 11.76 & 9.57 & 5.76 & 1.06 & 0.65 & 3.1 \\
RG(KDE0v) & CompOSE & 1.96 & 9.66 & 11.43 & 11.0 & 5.82 & 0.984 & 1.4 & 2.9 \\
RG(KDE0v1) & CompOSE & 1.97 & 9.79 & 11.64 & 10.8 & 5.53 & 0.965 & 0.92 & 1.9 \\
RG(Rs) & CompOSE & 2.12 & 10.76 & 12.95 & 8.96 & 4.23 & 0.922 & 2.8 & 3.7 \\
RG(Sk255) & CompOSE & 2.14 & 10.85 & 13.16 & 8.88 & 4.32 & 0.937 & 2.6 & 4.1 \\
RG(Sk272) & CompOSE & 2.23 & 11.09 & 13.33 & 8.40 & 4.30 & 0.965 & 2.8 & 5.0 \\
RG(Ska) & CompOSE & 2.21 & 10.89 & 12.92 & 8.63 & 4.46 & 0.970 & 2.4 & 4.4 \\
RG(Skb) & CompOSE & 2.19 & 10.61 & 12.22 & 8.85 & 4.64 & 0.974 & 1.6 & 3.8 \\
RG(SkI2) & CompOSE & 2.16 & 11.12 & 13.50 & 8.45 & 3.89 & 0.914 & 3.9 & 5.3 \\
RG(SkI3) & CompOSE & 2.24 & 11.31 & 13.57 & 8.08 & 3.89 & 0.937 & 3.8 & 5.5 \\
RG(SkI4) & CompOSE & 2.17 & 10.67 & 12.39 & 8.86 & 4.47 & 0.949 & 1.5 & 4.0 \\
RG(SkI5) & CompOSE & 2.24 & 11.47 & 14.10 & 7.96 & 3.76 & 0.932 & 5.0 & 7.0 \\
RG(SkI6) & CompOSE & 2.19 & 10.76 & 12.50 & 8.71 & 4.42 & 0.951 & 1.6 & 4.3 \\
RG(SkMp) & CompOSE & 2.11 & 10.53 & 12.51 & 9.26 & 4.58 & 0.946 & 1.9 & 2.8 \\
RG(SkOp) & CompOSE & 1.97 & 10.13 & 12.14 & 10.2 & 4.72 & 0.906 & 0.87 & 1.3 \\
RG(SLy2) & CompOSE & 2.05 & 10.05 & 11.80 & 10.1 & 5.34 & 0.977 & 0.39 & 1.5 \\
RG(SLy4) & CompOSE & 2.05 & 9.99 & 11.72 & 10.2 & 5.46 & 0.985 & 0.46 & 1.3 \\
RG(SLy9) & CompOSE & 2.16 & 10.63 & 12.48 & 9.01 & 4.59 & 0.952 & 1.3 & 3.9 \\
RG(SLy230a) & CompOSE & 2.10 & 10.25 & 11.85 & 9.60 & 4.95 & 0.948 & 1.3 & 4.7 \\
VGBCMR(D1M*) & CompOSE & 2.00 & 10.21 & 11.72 & 9.72 & 4.28 & 0.855 & 1.8 & 5.6 \\
XMLSLZ(DD-LZ1) & CompOSE & 2.56 & 12.28 & 13.14 & 6.38 & 3.12 & 0.875 & 1.6 & 2.3 \\
XMLSLZ(DDME-X) & CompOSE & 2.56 & 12.35 & 13.36 & 6.36 & 3.10 & 0.876 & 1.2 & 2.0 \\
XMLSLZ(GM1) & CompOSE & 2.36 & 11.94 & 13.75 & 7.08 & 3.17 & 0.851 & 2.1 & 6.3 \\
XMLSLZ(MTVTC) & CompOSE & 2.02 & 10.90 & 13.09 & 8.87 & 3.46 & 0.803 & 2.6 & 5.6 \\
XMLSLZ(NL3) & CompOSE & 2.77 & 13.28 & 14.58 & 5.53 & 2.81 & 0.890 & 1.2 & 1.7 \\
XMLSLZ(PK1) & CompOSE & 2.31 & 12.66 & 14.36 & 6.33 & 2.14 & 0.682 & 1.7 & 2.5 \\
XMLSLZ(PKDD) & CompOSE & 2.33 & 11.77 & 13.62 & 7.33 & 3.30 & 0.852 & 1.6 & 5.7 \\
XMLSLZ(TM1) & CompOSE & 2.18 & 12.37 & 14.27 & 6.72 & 2.11 & 0.659 & 1.4 & 4.3 \\
XMLSLZ(TW99) & CompOSE & 2.08 & 10.62 & 12.26 & 9.06 & 3.96 & 0.839 & 1.8 & 6.2 \\
YHSHP(PAL1\_120) & Y+11 & 1.47 & 9.18 & 10.37 & 13.8 & 4.09 & 0.699 & 2.1 & 7.4 \\
YHSHP(PAL1\_180) & Y+11 & 1.74 & 9.93 & 12.04 & 11.2 & 3.97 & 0.771 & 1.1 & 1.6 \\
YHSHP(PAL1\_240) & Y+11 & 1.95 & 10.60 & 12.77 & 9.50 & 3.70 & 0.811 & 1.9 & 3.7 \\
YHSHP(PAL2\_120) & Y+11 & 1.48 & 9.72 & 11.33 & 12.6 & 3.47 & 0.677 & 2.8 & 4.0 \\
YHSHP(PAL2\_180) & Y+11 & 1.75 & 10.35 & 12.93 & 10.5 & 3.52 & 0.753 & 3.2 & 4.4 \\
YHSHP(PAL2\_240) & Y+11 & 1.97 & 10.97 & 13.55 & 9.03 & 3.35 & 0.795 & 3.5 & 5.4 \\
YHSHP(PAL2\_400\_0) & Y+11 & 2.85 & 12.69 & 14.39 & 6.00 & 4.49 & 1.23 & 4.5 & 8.4 \\
YHSHP(PAL3\_120) & Y+11 & 1.42 & 8.45 & 8.95 & 15.9 & 5.14 & 0.729 & 5.8 & 14 \\
YHSHP(PAL3\_180) & Y+11 & 1.69 & 9.36 & 11.07 & 12.3 & 4.66 & 0.799 & 1.8 & 3.7 \\
YHSHP(PAL3\_240) & Y+11 & 1.91 & 10.12 & 11.94 & 10.2 & 4.18 & 0.828 & 0.45 & 2.2 \\
YHSHP(PAL3\_260\_01) & Y+11 & 2.06 & 10.41 & 12.20 & 9.47 & 4.46 & 0.907 & 0.86 & 3.0 \\
YHSHP(PAL3\_300) & Y+11 & 1.83 & 10.51 & 12.22 & 9.47 & 3.02 & 0.709 & 1.0 & 4.0 \\
YHSHP(PAL3\_300\_0) & Y+11 & 2.34 & 11.08 & 12.71 & 8.07 & 4.69 & 1.04 & 2.1 & 4.9 \\
YHSHP(PAL3\_400) & Y+11 & 2.83 & 12.32 & 13.30 & 6.20 & 4.92 & 1.26 & 3.0 & 6.7 \\
YHSHP(PAPAL\_120) & Y+11 & 1.44 & 8.74 & 9.55 & 15.0 & 4.70 & 0.717 & 4.2 & 12 \\
YHSHP(PAPAL\_180) & Y+11 & 1.71 & 9.59 & 11.47 & 11.9 & 4.37 & 0.786 & 1.0 & 2.4 \\
YHSHP(PAPAL\_240) & Y+11 & 1.93 & 10.32 & 12.29 & 9.92 & 3.98 & 0.822 & 0.94 & 2.7 \\
    \hline
    \multicolumn{10}{c}{Models with hyperons and  $\Delta$-isobars}\\
    G(GNH3) & R+09 & 1.96 & 11.39 & 14.18 & 8.49 & 2.85 & 0.747 & 4.7 & 8.3 \\
LNO(H1) & R+09 & 1.55 & 10.97 & 12.85 & 9.16 & 1.95 & 0.541 & 4.7 & 7.3 \\
LNO(H2) & R+09 & 1.67 & 11.51 & 13.47 & 8.02 & 1.71 & 0.564 & 4.5 & 6.8 \\
LNO(H3) & R+09 & 1.79 & 11.85 & 13.82 & 7.52 & 1.68 & 0.518 & 3.0 & 6.3 \\
LNO(H4) & R+09 & 2.03 & 11.76 & 13.74 & 7.51 & 2.30 & 0.653 & 3.4 & 6.7 \\
LNO(H5) & R+09 & 1.73 & 11.31 & 13.23 & 8.19 & 1.92 & 0.596 & 4.4 & 6.2 \\
LNO(H7) & R+09 & 1.68 & 10.85 & 12.92 & 9.18 & 2.32 & 0.630 & 4.4 & 6.3 \\
RSGMT(QMC700) & OF16 & 1.98 & 12.36 & 12.84 & 5.99 & 1.17 & 0.404 & 21 & 34 \\
BG(BGN1H1) & R+09 & 1.63 & 9.33 & 12.89 & 13.6 & 5.48 & 0.876 & 11 & 55 \\
BG(BGN1H2) & HPY07 & 1.57 & 9.04 & 12.52 & 14.6 & 5.95 & 0.897 & 14 & 62 \\
BG(BGN2H1) & HPY07 & 1.82 & 9.57 & 13.52 & 12.4 & 6.36 & 1.01 & 7.4 & 31 \\
BG(BGN2H2) & HPY07 & 1.74 & 9.12 & 13.15 & 13.9 & 7.45 & 1.05 & 11 & 49 \\
DNS(CMF) & CompOSE & 2.07 & 11.88 & 13.59 & 7.19 & 2.19 & 0.686 & 2.1 & 4.8 \\
DS(CMF)-1 & CompOSE & 2.07 & 11.88 & 13.59 & 7.20 & 2.19 & 0.686 & 2.1 & 4.8 \\
DS(CMF)-2 & CompOSE & 2.13 & 12.00 & 13.71 & 7.05 & 2.25 & 0.692 & 2.0 & 5.6 \\
DS(CMF)-3 & CompOSE & 2.00 & 11.54 & 13.19 & 7.65 & 2.32 & 0.685 & 1.5 & 4.4 \\
DS(CMF)-4 & CompOSE & 2.05 & 11.59 & 13.28 & 7.59 & 2.41 & 0.689 & 1.6 & 5.3 \\
DS(CMF)-5 & CompOSE & 2.07 & 11.48 & 13.23 & 7.74 & 2.65 & 0.753 & 2.4 & 4.9 \\
DS(CMF)-6 & CompOSE & 2.11 & 11.57 & 13.32 & 7.62 & 2.65 & 0.744 & 2.1 & 5.5 \\
DS(CMF)-7 & CompOSE & 2.07 & 11.49 & 13.23 & 7.73 & 2.64 & 0.752 & 2.4 & 4.9 \\
DS(CMF)-8 & CompOSE & 2.09 & 11.58 & 13.32 & 7.60 & 2.58 & 0.744 & 2.4 & 4.9 \\
GHK(GM1A) & O+19 & 1.99 & 11.94 & 13.72 & 7.18 & 2.02 & 0.686 & 5.7 & 7.9 \\
GHK(TM1C) & O+19 & 2.05 & 12.47 & 14.31 & 6.58 & 1.77 & 0.644 & 3.0 & 4.1 \\
GHK(GM1'B) & GHK14 & 2.01 & 11.46 & 13.76 & 8.13 & 2.78 & 0.748 & 5.0 & 7.4 \\
OGHF(FSU2H) & O+19 & 1.99 & 11.98 & 13.31 & 6.90 & 1.83 & 0.637 & 3.4 & 4.8 \\
OGHF(NL3$\omega\rho$) & O+19 & 2.71 & 12.94 & 13.73 & 5.73 & 2.85 & 0.889 & 0.67 & 0.96 \\
OPGR(GM1Y4) & CompOSE & 1.79 & 13.01 & 13.77 & 5.23 & 0.823 & 0.424 & 1.5 & 3.3 \\
OPGR(GM1Y5) & CompOSE & 2.12 & 12.30 & 13.77 & 6.61 & 1.91 & 0.667 & 3.2 & 5.2 \\
OPGR(GM1Y6) & CompOSE & 2.29 & 12.12 & 13.77 & 6.84 & 2.64 & 0.808 & 3.0 & 4.1 \\
R(DD2Y$\Delta$)1.1-1.1 & CompOSE & 2.04 & 11.21 & 13.00 & 8.22 & 2.89 & 0.741 & 1.5 & 6.2 \\
R(DD2Y$\Delta$)1.2-1.1 & CompOSE & 2.05 & 10.98 & 12.31 & 8.39 & 3.05 & 0.731 & 3.4 & 5.9 \\
R(DD2Y$\Delta$)1.2-1.3 & CompOSE & 2.03 & 11.44 & 13.26 & 7.89 & 2.63 & 0.739 & 3.8 & 5.3 \\
    \hline
    \multicolumn{10}{c}{Hybrid models}\\
    ABPR(ALF1) & R+09 & 1.49 & 9.21 & 9.89 & 11.9 & 2.95 & 0.565 & 8.9 & 26 \\
ABPR(ALF2) & R+09 & 2.09 & 11.96 & 13.18 & 6.94 & 1.95 & 0.553 & 4.3 & 6.5 \\
ABPR(ALF3) & R+09 & 1.47 & 9.50 & 10.31 & 11.6 & 2.75 & 0.565 & 5.4 & 15 \\
ABPR(ALF4) & R+09 & 1.94 & 10.88 & 11.66 & 8.28 & 2.33 & 0.506 & 13 & 20 \\
PCL(PCL2) & R+09 & 1.48 & 10.18 & 11.73 & 10.8 & 2.54 & 0.599 & 5.2 & 8.5 \\
BFH(QHC19-A) & CompOSE & 1.93 & 10.23 & 11.55 & 9.63 & 3.66 & 0.781 & 2.1 & 5.3 \\
BFH(QHC19-B) & CompOSE & 2.07 & 10.55 & 11.59 & 8.82 & 3.59 & 0.799 & 4.5 & 11 \\
BFH(QHC19-C) & CompOSE & 2.18 & 10.74 & 11.59 & 8.37 & 3.70 & 0.815 & 7.7 & 18 \\
BFH(QHC19-D) & CompOSE & 2.28 & 10.86 & 11.59 & 8.12 & 3.92 & 0.833 & 11 & 23 \\
BFH(QHC18) & CompOSE & 2.04 & 10.38 & 11.48 & 9.22 & 3.88 & 0.804 & 7.2 & 21 \\
DS(CMF)-1 & CompOSE & 1.97 & 11.19 & 13.59 & 8.54 & 2.98 & 0.784 & 9.9 & 20 \\
DS(CMF)-2 & CompOSE & 1.96 & 11.14 & 13.71 & 8.70 & 3.08 & 0.787 & 10 & 22 \\
DS(CMF)-3 & CompOSE & 1.99 & 11.24 & 13.19 & 8.21 & 2.78 & 0.777 & 6.7 & 14 \\
DS(CMF)-4 & CompOSE & 1.98 & 11.24 & 13.28 & 8.27 & 2.82 & 0.778 & 8.2 & 17 \\
DS(CMF)-5 & CompOSE & 2.02 & 11.88 & 13.24 & 6.92 & 1.93 & 0.727 & 4.7 & 9.8 \\
DS(CMF)-6 & CompOSE & 2.01 & 11.99 & 13.32 & 6.76 & 1.84 & 0.728 & 6.6 & 14 \\
DS(CMF)-7 & CompOSE & 2.02 & 11.89 & 13.24 & 6.92 & 1.92 & 0.727 & 4.9 & 11 \\
DS(CMF)-8 & CompOSE & 2.02 & 11.89 & 13.32 & 6.92 & 1.92 & 0.727 & 6.0 & 13 \\
JJ(VQCD)intermediate & CompOSE & 2.15 & 11.80 & 12.41 & 6.12 & 1.92 & 0.689 & 4.7 & 12 \\
JJ(VQCD)soft & CompOSE & 2.03 & 11.85 & 12.33 & 5.48 & 1.48 & 0.648 & 7.7 & 17 \\
JJ(VQCD)stiff & CompOSE & 2.33 & 11.86 & 12.54 & 6.81 & 2.68 & 0.763 & 2.4 & 3.6 \\
KBH(QHC21\_A) & CompOSE & 2.18 & 11.24 & 12.37 & 7.83 & 3.15 & 0.798 & 1.2 & 3.8 \\
KBH(QHC21\_AT) & CompOSE & 2.13 & 10.87 & 11.84 & 8.25 & 3.35 & 0.798 & 2.0 & 4.0 \\
KBH(QHC21\_B) & CompOSE & 2.25 & 11.41 & 12.36 & 7.49 & 3.08 & 0.803 & 2.3 & 5.9 \\
KBH(QHC21\_BT) & CompOSE & 2.20 & 11.07 & 11.85 & 7.86 & 3.25 & 0.803 & 3.0 & 6.3 \\
KBH(QHC21\_C) & CompOSE & 2.31 & 11.56 & 12.36 & 7.22 & 3.03 & 0.808 & 3.6 & 8.3 \\
KBH(QHC21\_CT) & CompOSE & 2.26 & 11.13 & 11.74 & 7.68 & 3.31 & 0.812 & 5.5 & 10 \\
KBH(QHC21\_D) & CompOSE & 2.36 & 11.69 & 12.34 & 6.99 & 3.01 & 0.813 & 5.0 & 11 \\
KBH(QHC21\_DT) & CompOSE & 2.32 & 11.28 & 11.75 & 7.41 & 3.27 & 0.817 & 6.8 & 13 \\
OOS(DD2\_FRG)(2) & CompOSE & 2.05 & 12.53 & 13.22 & 5.92 & 1.36 & 0.551 & 5.2 & 16 \\
OOS(DD2\_FRG)(2+1) & CompOSE & 1.84 & 12.75 & 13.22 & 5.33 & 0.897 & 0.466 & 4.9 & 15 \\
OOS(DD2-FRG)vec(2) & CompOSE & 2.12 & 12.69 & 13.22 & 5.67 & 1.32 & 0.559 & 12 & 27 \\
OOS(DD2-FRG)vec(2+1) & CompOSE & 1.93 & 12.98 & 13.22 & 4.99 & 0.844 & 0.467 & 14 & 34 
\end{longtable}
}

\end{document}